\documentclass[twocolumn,showpacs,showketwd,prd,preprintnumbers,amsmath,amssymb]{revtex4}

\usepackage{graphicx}
\usepackage{dcolumn}
\usepackage{bm}
\usepackage{epsfig}
\usepackage{amssymb}
\usepackage{amsmath}

 \def\tskip{\setlength{\tskip}{5pt}}
\def\colwidth{\setlength{\colwidth}{3.5in}}

\def\prd{Phys. Rev. D}

\def\prl{Phys. Rev. Lett.}

\def\apj{Astrophys. J.}
\def\apjl{Astrophys. J. Lett.}

\def\mnras{Mon. Not. Roy. Astron. Soc.}
\def\apjs{Astrophys. J. Suppl. Ser.}

\newcommand{\lsim}{\mathrel{\hbox{\rlap{\lower.55ex\hbox{$\sim$}} \kern-.3em \raise.4ex \hbox{$<$}}}}
\newcommand{\gsim}{\mathrel{\hbox{\rlap{\lower.55ex\hbox{$\sim$}} \kern-.3em \raise.4ex \hbox{$>$}}}}
\newcommand{\beq}{\begin{equation}}
\newcommand{\eeq}{\end{equation}}
\newcommand{\beqa}{\begin{eqnarray}}
\newcommand{\eeqa}{\end{eqnarray}}

\begin{document}

\title{Anisotropy of Cosmic Acceleration}

\author{Wen Zhao$^{1}$,  Puxun Wu$^{2}$ and Yang Zhang$^{1}$}
\affiliation{
$^{1}$Key Laboratory for Researches in Galaxies and Cosmology, Department of Astronomy, University of Science and Technology of China, Hefei, Anhui, 230026, China\\
$^{2}$Center for Nonlinear Science and Department of Physics, Ningbo University, Ningbo, Zhejiang, 315211, China}

\date{\today}

\begin{abstract}

In this paper, we study the anisotropy of cosmic acceleration by
dividing the Union2 Type Ia supernova dataset into 12 subsets
according to their positions in Galactic coordinate system. In
each region, we derive the deceleration parameter $q_0$ as the
diagnostic to quantify the anisotropy level in the corresponding
direction, and construct $q_0$ anisotropic maps by combining these
$q_0$ values. In addition to the monopole component, we find the
significant dipole effect in the $q_0$-maps with the amplitude
$A_1=0.466^{+0.255}_{-0.205}$, which deviates from zero at more
than 2-$\sigma$ level. The direction of the best-fit dipole is
($\theta=108.8^{\circ}$, $\phi=187.0^{\circ}$) in Galactic system.
Interesting enough, we find the direction of this dipole is nearly
perpendicular to the CMB kinematic dipole, and the angle between
them is $95.7^{\circ}$. The perpendicular relation is anomalous at
the 1-in-10 level.

\end{abstract}


\maketitle



\section{Introduction \label{section-1}}

Soon after the discovery of the accelerating cosmic expansion from
the observations of Type Ia supernova (SNIa) \cite{snia1, snia2},
a number of authors have investigated the anisotropies of the
cosmic acceleration
\cite{peculiar,peculiar2,peculiar3,peculiar4,fluctuation,fluctuation2,fluctuation3,others,others2,others3,others4,hemi1,hemi2,hemi22,cai,kalus,cai2},
which was motivated in several aspects: From the theoretical point
of view, the anisotropy may arise in some cosmological models,
such as the vector dark energy models
\cite{vector,ym,ym2,ym3,vector2}, the anisotropic
equation-of-state of dark energy\cite{ani}, the non-trivial cosmic
topology \cite{bianchi,bianchi2,topology}, the statistically
anisotropic primordial perturbations
\cite{perturbation,perturbation2} or the existence of a
large-scale primordial magnetic field
\cite{magnetic,magnetic2,magnetic3}. And also the study can be
used to check the validity of the cosmological principle
\cite{principle,principle2,principle3}, to measure expected
deviations from the isotropy due to various motions of Local Group
\cite{peculiar,peculiar2,peculiar3,peculiar4}, or to search for
the possible systematic errors in observations and their analysis
\cite{hemi1}.

In particular, several groups \cite{hemi1,hemi2,hemi22,cai,cai2}
have applied the hemisphere comparison method to study the
anisotropy of $\Lambda$CDM, $w$CDM and the dark energy model with
CPL parametrization, where the supernova data and the
corresponding cosmic accelerations on several pairs of opposite
hemispheres have been used to search for maximally asymmetric
pair, and a statistically significant preferred axis has been
reported. As emphasized in \cite{hemi2,hemi22}, although this
method optimizes the statistics due to the large number of
supernovae in each hemisphere, it has lost all information about
the detailed structure of the anisotropy.

In this paper, we shall extend this issue to investigate the cosmic acceleration in different parts of the whole sky, and study the possible existence of anisotropy. To do it,
we take use of the Union2 dataset \cite{union2}, and divide them into 12 parts according to their positions in Galactic coordinate system. Among them, six regions are useless and masked in the investigation due to the lack of supernova data. In each unmasked region, we study the cosmic acceleration by taking the deceleration parameter $q_0$ as the diagnostic, and find the significant difference for different regions. We extract the lowest multipole components, i.e. monopole and dipole, in this anisotropic map, and find that the monopole amplitude is $A_0=-0.750^{+0.122}_{-0.172}$, which is consistent with other observations, and shows the present acceleration of cosmic expansion. Meanwhile, we find the significant dipole, and the amplitude is $A_1=0.466^{+0.255}_{-0.205}$, which deviates from zero in more than 2-$\sigma$ confident level. Interesting enough, the direction of the best-fit dipole is at ($\theta=108.8^{\circ}$, $\phi=187.0^{\circ}$) \footnote{Throughout this paper, we use the polar coordinate ($\theta$,$\phi$) in the Galactic system, which relates to the
Galactic coordinate ($l$, $b$) by $l = 90^{\circ}-\theta$ and $b=\phi$.}, which is nearly perpendicular to CMB kinematic dipole, and the angle between these two dipole directions is $95.7^{\circ}$. This implies that the origin of the anisotropy of cosmic acceleration may connect with the CMB kinematic dipole.

The paper is organized as follows. In the next section we give a general introduction to the analysis method, and apply to $w$CDM model fitted by the Union2 dataset. In this section, we focus on the monopole and dipole components, especially the direction of dipole, and compare with CMB kinematic dipole. Sec. 3 gives our conclusions.

\section{Anisotropy of Cosmic Acceleration \label{section-2}}

\begin{figure}[t]
\begin{center}
\includegraphics[width = 4cm]{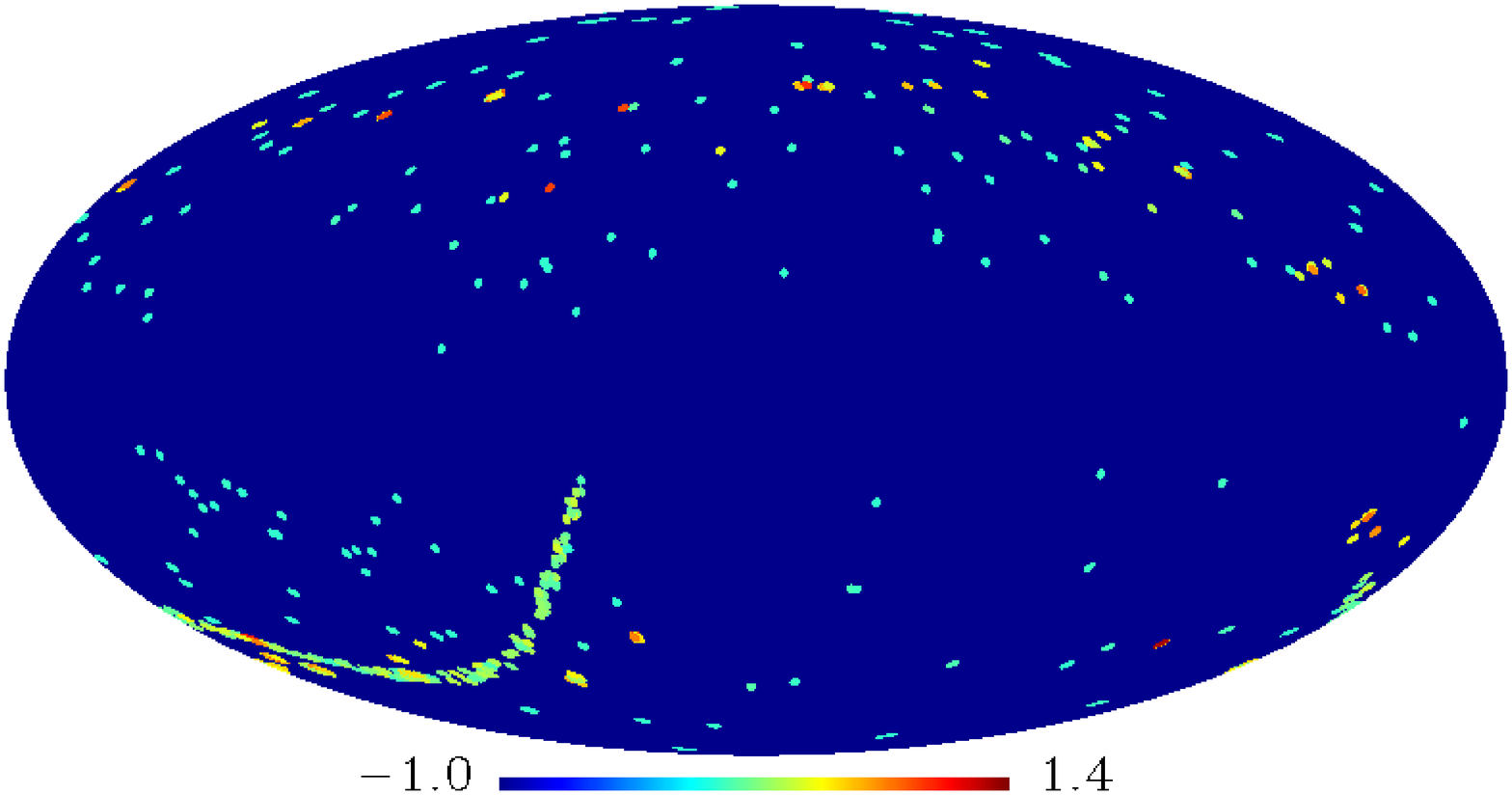}\includegraphics[width = 4cm]{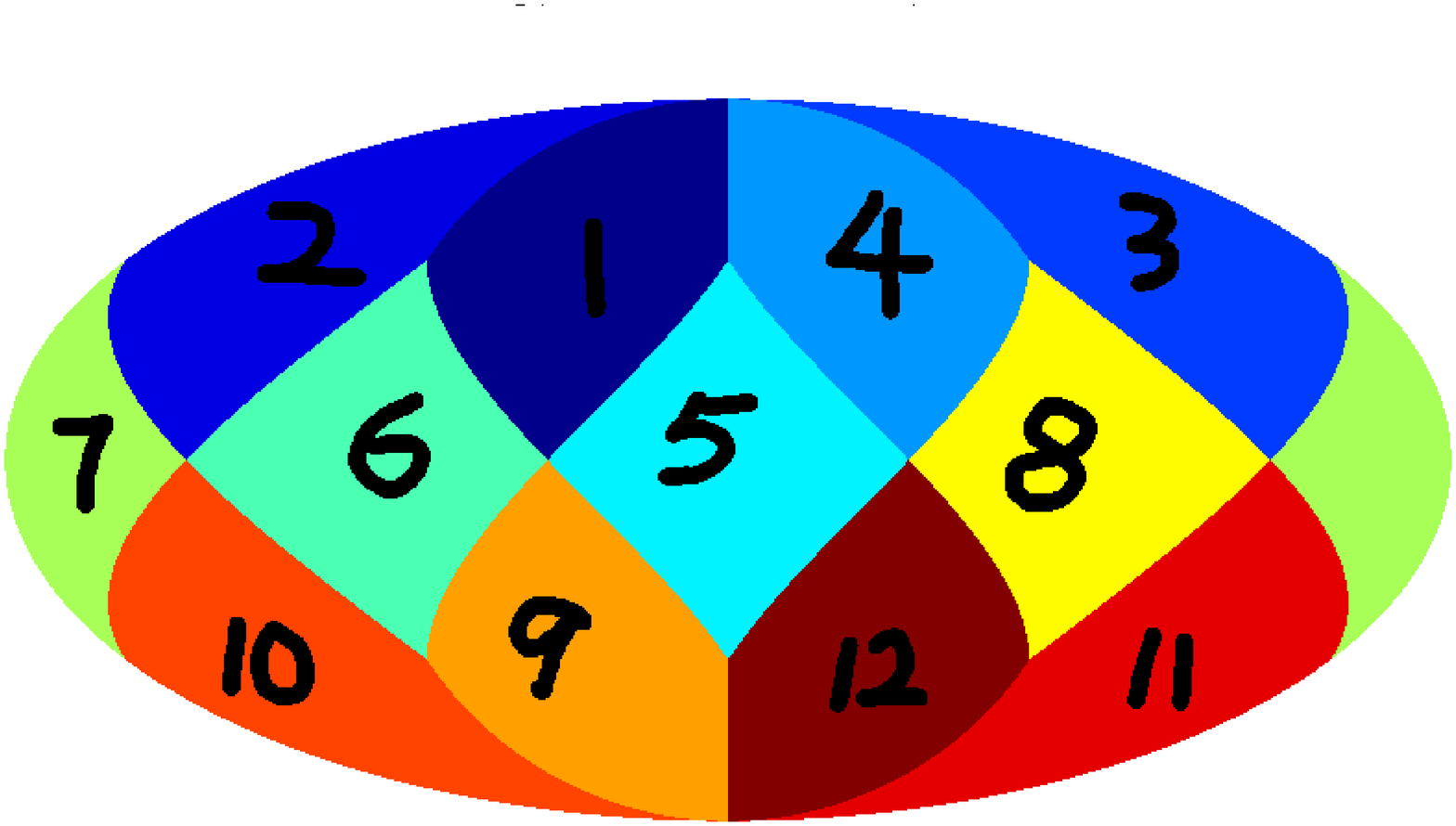}
\end{center}
\caption{Left Panel: The distribution of the SNIa in Galactic coordinate system, where the color indicates the redshift of SNIa.
Right Panel: The whole sky is divided into 12 equal-area regions. }\label{figure1}
\end{figure}

In this paper we take use of the Union2 dataset \cite{union2},
which contains 557 type Ia SNIa data and uses SALT2 for SNIa
light-curve fitting, covering the redshift range  $z=[0.015, 1.4]$
and including samples from other surveys, such as CfA3 \cite{cfa},
SDSS-II Supernova Search \cite{sdss} and high-z Hubble Space
Telescope. The direction distribution of the supernovae in
Galactic coordinate system is presented in Fig. \ref{figure1}
(left panel) \cite{fluctuation,fluctuation2,fluctuation3}.


In order to investigate the cosmic acceleration in different directions, in principle we can divide the total supernovae into a number of groups according to their positions and redshifts. Limited by the total number of the SNIa, in this paper we shall ignore the possible redshift effect of the anisotropy. HEALPix is a genuinely curvilinear partition of the 2-dimensional sphere into exactly equal area quadrilaterals of varying shape, which is advantageous since sky signals are sampled without regional dependence \cite{healpix}. The base-resolution comprises 12 pixels in three ring around the poles and equator. The resolution of the grid is expressed by the parameter Nside. In this paper, we adopt the lowest resolution with Nside=1 due to the fact that the total number of SNIa is not too large. The 12 regions are shown in Fig. \ref{figure1} (right panel).
We find that the distribution of SNIa in the sky is not isotropic. In Regions 5, 6, 7, 8, i.e. Galactic plane, the numbers of SNIa are very smaller (i.e. $N_{\rm sn}=1$ in Region 5, $N_{\rm sn}=7$ in Region 6, $N_{\rm sn}=11$ in Region 7, $N_{\rm sn}=4$ in Region 11). In addition, we also find that $N_{\rm sn}=22$ in Region 1, and $N_{\rm sn}=5$ in Region 12. So in the following discussion, we will not use these six regions, because of the small numbers of SNIa. In Table \ref{table1}, we list the values of $N_{\rm sn}$ in the other six regions, which will be used for the analysis.

\begin{table*}
\caption{Results for $N_{\rm sn}$, $\Omega_m$, $w$ and $q_0$ in the six unmasked regions.}
\begin{center}
\label{table1}
\begin{tabular}{|c|c|c|c|c|c|c|c|}
         \hline
 & Region 2& Region 3& Region 4& Region 9& Region 10& Region 11\\
          \hline
 $N_{\rm sn}$ & $82$ &$62$& $35$& $104$ &$181$& $43$ \\
          \hline
 $\Omega_m$ & $0.330^{+0.093}_{-0.150}$ &$0.000^{+0.289}$& $0.301^{+0.148}_{-0.301}$& $0.417^{+0.104}_{-0.284}$ &$0.133^{+0.226}_{-0.133}$& $0.152^{+0.245}_{-0.152}$ \\
           \hline
 $w$ & $-1.270^{+0.453}_{-0.610}$ &$-0.755^{+0.095}_{-0.650}$& $-1.291^{+0.653}_{-1.191}$& $-1.423^{+0.756}_{-0.949}$ &$-0.742^{+0.193}_{-0.483}$& $-0.662^{+0.222}_{-0.659}$\\
  \hline
 $q_0$ & $-0.777^{+0.278}_{-0.390}$ &$-0.616^{+0.328}_{-0.302}$& $-0.853^{+0.396}_{-0.740}$& $-0.745^{+0.393}_{-0.544}$ &$-0.464^{+0.140}_{-0.239}$& $-0.342^{+0.273}_{-0.390}$\\
  \hline
\end{tabular}
\end{center}
\end{table*}

In each region, we fit the SNIa data by minimizing the $\chi_{\rm sn}^2$ values of the distance modulus. The $\chi^2_{\rm sn}$ for SNIa is obtained by comparing theoretical distance modulus $\mu_{th}(z)=5\log_{10}[d_L(z)]+\mu_0$, where $\mu_0=42.384-5\log_{10}h$ is a nuisance parameter, with observed $\mu_{ob}$ of supernovae:
 \begin{eqnarray}
 \chi_{\rm sn}^2=\sum_{i=1}^{N_{\rm sn}}\frac{[\mu_{th}(z_i)-\mu_{ob}(z_i)]^2}{\sigma^2(z_i)}.
 \end{eqnarray}
The expressions of $d_L(z)$ and $H(z)$ depend on the cosmological model. In this paper we consider the flat Friedmann-Lemaitre-Robertson-Walker Universe and the $w$CDM model. So one has
 \begin{eqnarray}
 d_L(z)=(1+z)\int_0^z \frac{H_0}{H(z')}dz',
 \end{eqnarray}
and
 \begin{equation}
 H^2(z)=H_0^2[\Omega_{m}(1+z)^2+(1-\Omega_m)(1+z)^{3+3w}].
 \end{equation}

The nuisance parameter $\mu_0$ can be eliminated in the following way. We expand $\chi^2_{\rm sn}$ with
respect to $\mu_0$ \cite{chi2}:
 \begin{equation}
 \chi^2_{\rm sn}=C_2+2C_1\mu_0+C_0\mu_0^2, \label{26}
 \end{equation}
where
 \[
 C_k=\sum_i\frac{[\mu_{th}(z_i;\mu_0=0)-\mu_{ob}(z_i)]^k}{\sigma^2(z_i)},~~~(k=0,1,2).
 \]
Eq. (\ref{26}) has a minimum as follows,
 \begin{equation}
 {\tilde{\chi}_{\rm sn}}^2=\chi^2_{{\rm sn},min}=C_2-C_1^2/C_0,
 \end{equation}
which is independent of $\mu_0$. Actually, the difference between ${\tilde{\chi}_{\rm sn}}^2$ and the marginalized $\chi^2_{\rm sn}$ is just a constant \cite{chi2}. In the following analysis, we will adopt ${\tilde{\chi}_{\rm sn}}^2$ as the goodness of fitting between theoretical model and SNIa data.

Similar to \cite{cai}, we use the deceleration parameter $q_0$ as the diagnostic of the cosmic acceleration. In the $w$CDM model, the present value of $q_0$ can be expressed as
 \begin{equation}
 q_0=\frac{1+3w(1-\Omega_m)}{2},
 \end{equation}
which is a combination of the parameters $w$ and $\Omega_m$. The results of the cosmological parameters $\Omega_m$, $w$ and the corresponding $q_0$ in each region are shown in Table \ref{table1}. Interesting, we find the significant difference for the different regions, even if the error bars are considered. For example, the absolute values of $q_0$ are quite small in Regions 10 and 11, but fairly large in Regions 2, 4, 9. The similar difference also exists for the parameters $\Omega_m$ and $w$. These are the clear indications of the anisotropy of the cosmic acceleration.

\begin{figure}[t]
\begin{center}
\includegraphics[width = 5cm]{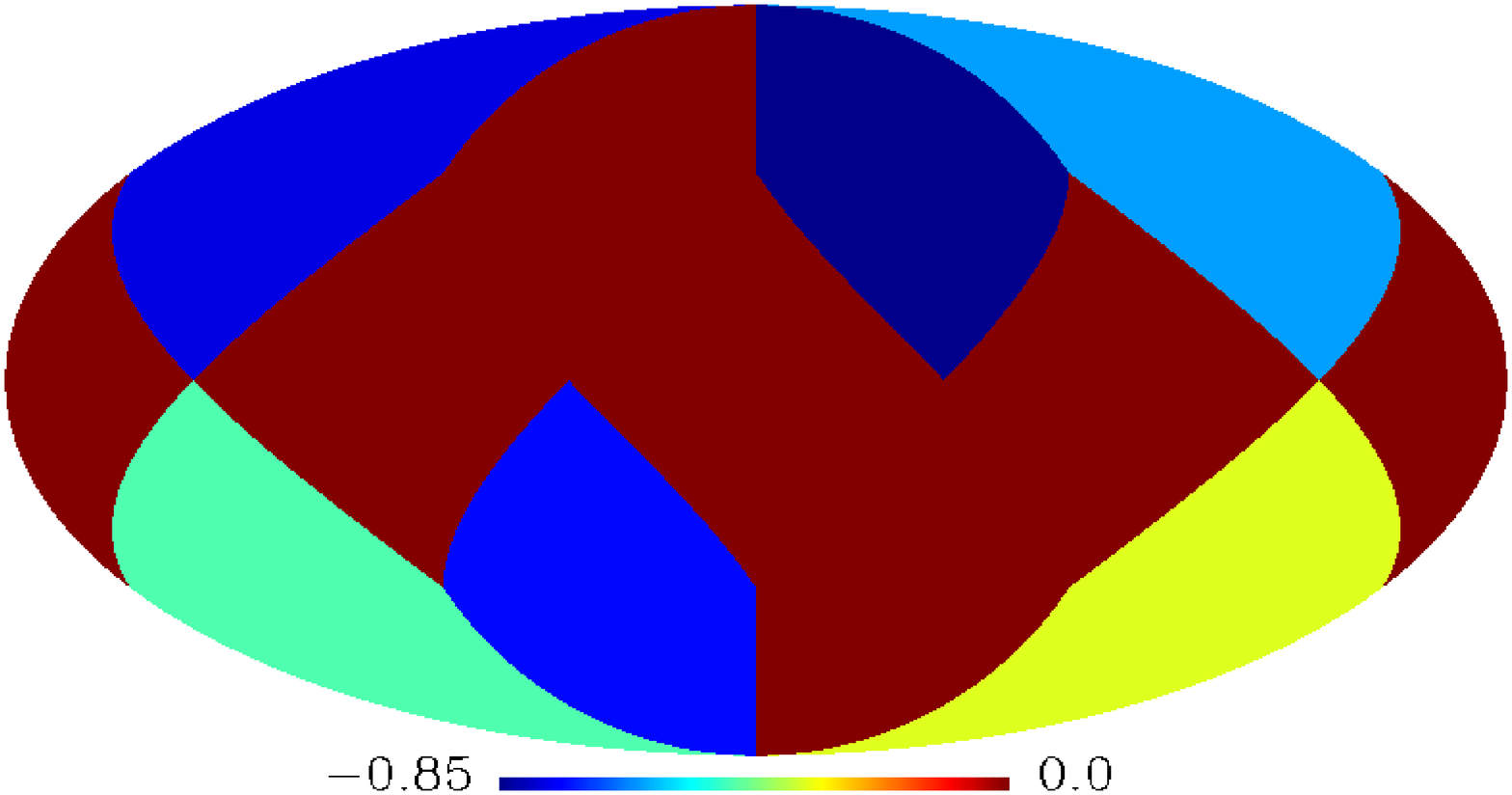} \\
\includegraphics[width = 5cm]{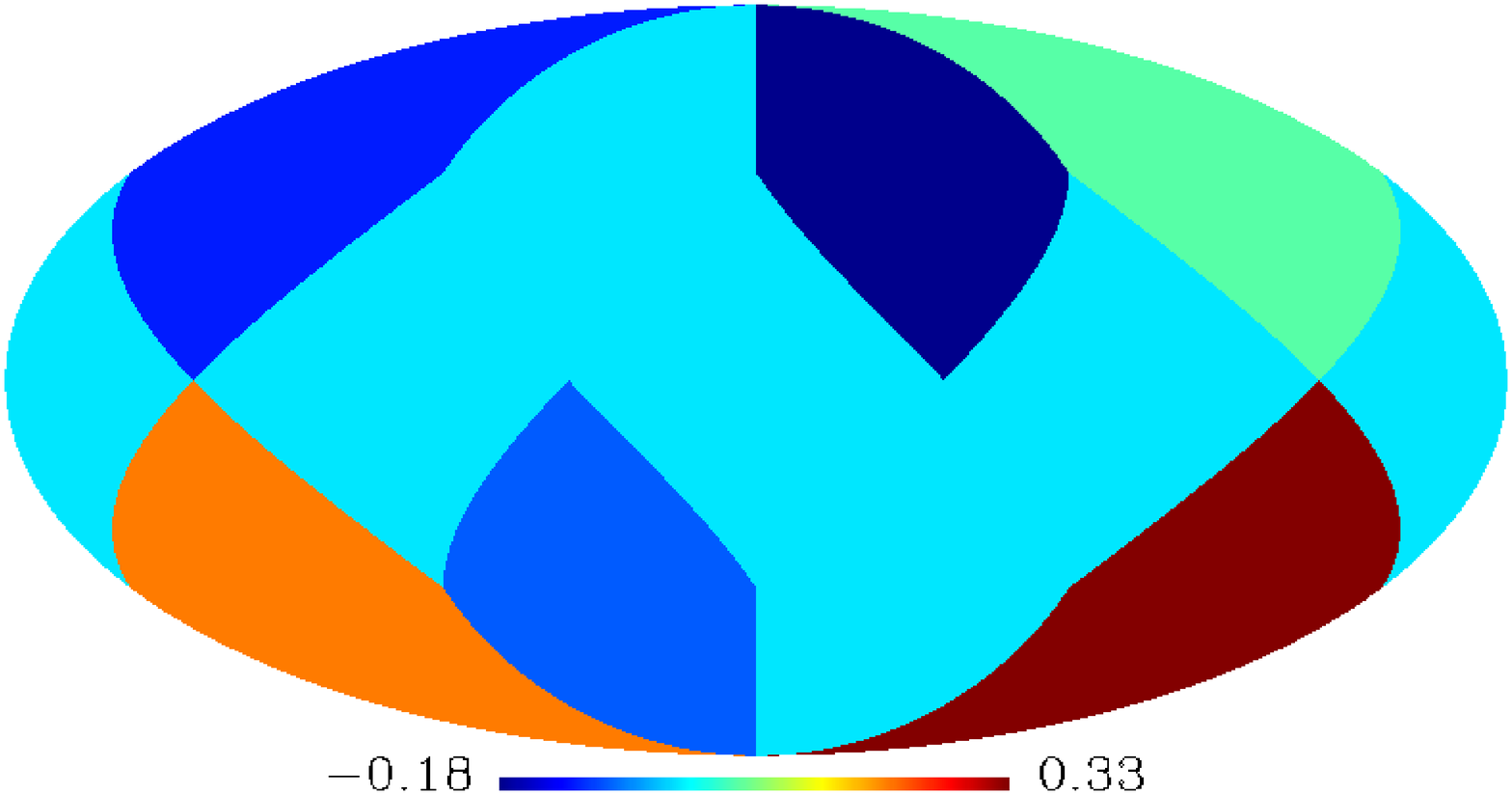}\\
\includegraphics[width = 5cm]{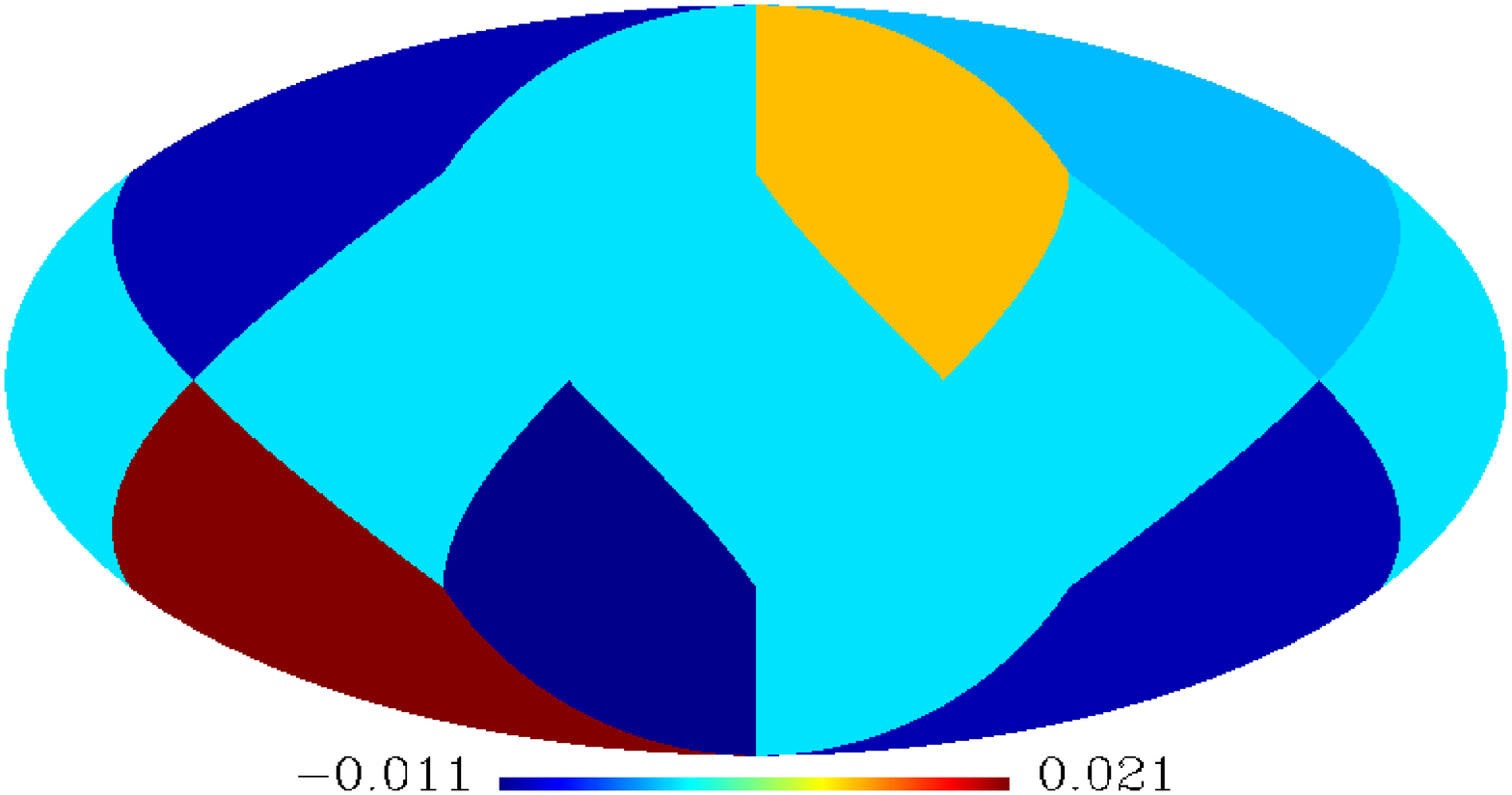}
\end{center}
\caption{Upper Panel: The best-fit $q_0$ values in different regions. Middle Panel: The monopole component is subtracted in upper panel. Lower Panel: Both monopole and dipole are subtracted in upper panel.}\label{figure2}
\end{figure}

As the first step to quantify the anisotropy, we use the best-fit
$q_0$ values as the diagnostic, which is shown in Fig.
\ref{figure2} (upper panel). In order to describe the
2-dimensional anisotropic map, it is convenient to expand it over
the spherical harmonics as follows:
 \begin{equation}
 q_0(\theta,\phi)=\sum_{l=0}^{l_{\max}}\sum_{m=-l}^{l} a_{lm} Y_{lm}(\theta,\phi),
 \end{equation}
where $Y_{lm}$ are the spherical harmonics, and $a_{lm}$ are the
corresponding multipole coefficients. In this paper, due to the
low resolution of the anisotropic map, we only consider the lowest
multipoles: monopole with $l=0$ and dipole with $l=1$.

We apply the routine provided in HEALPix package to subtract the
monopole and dipole components from the partial HEALPix map
\cite{healpix}. The fit is obtained by solving the linear system
 \begin{equation}
 \sum_{j=0}^{3}A_{ij} f_j=b_i,
 \end{equation}
and
 \begin{equation}
 b_i=\sum_{p\in{\mathcal P}} s_i(p) m(p),~~~~
 A_{ij}=\sum_{p\in{\mathcal P}} s_i(p) s_j(p),
 \end{equation}
where $\mathcal{P}$ is the set of valid, unmasked pixels, and $m(p)$ is the input map. $s_0(p)=1$ and $s_1(p)=x$, $s_2(p)=y$, $s_3(p)=z$ are respectively the monopole and dipole templates. The output monopole and dipole are respectively,
 \begin{equation}
 m_{\rm monopole}(p)=f_0,~~~~
 m_{\rm dipole}(p)=\sum_{i=1}^{3}f_i s_i(p).
 \end{equation}

Applying to the $q_0$-map in Fig. \ref{figure2} (upper panel), we
obtain the fit monopole $A_0=-0.674$, which is equivalent to the
average decelerating parameter in the whole sky. This value
clearly shows that the present Universe is in an accelerating
expansion stage, which is consistent with other results
\cite{other_results,other_results2}.

\begin{figure}[t]
\begin{center}
\includegraphics[width = 4cm]{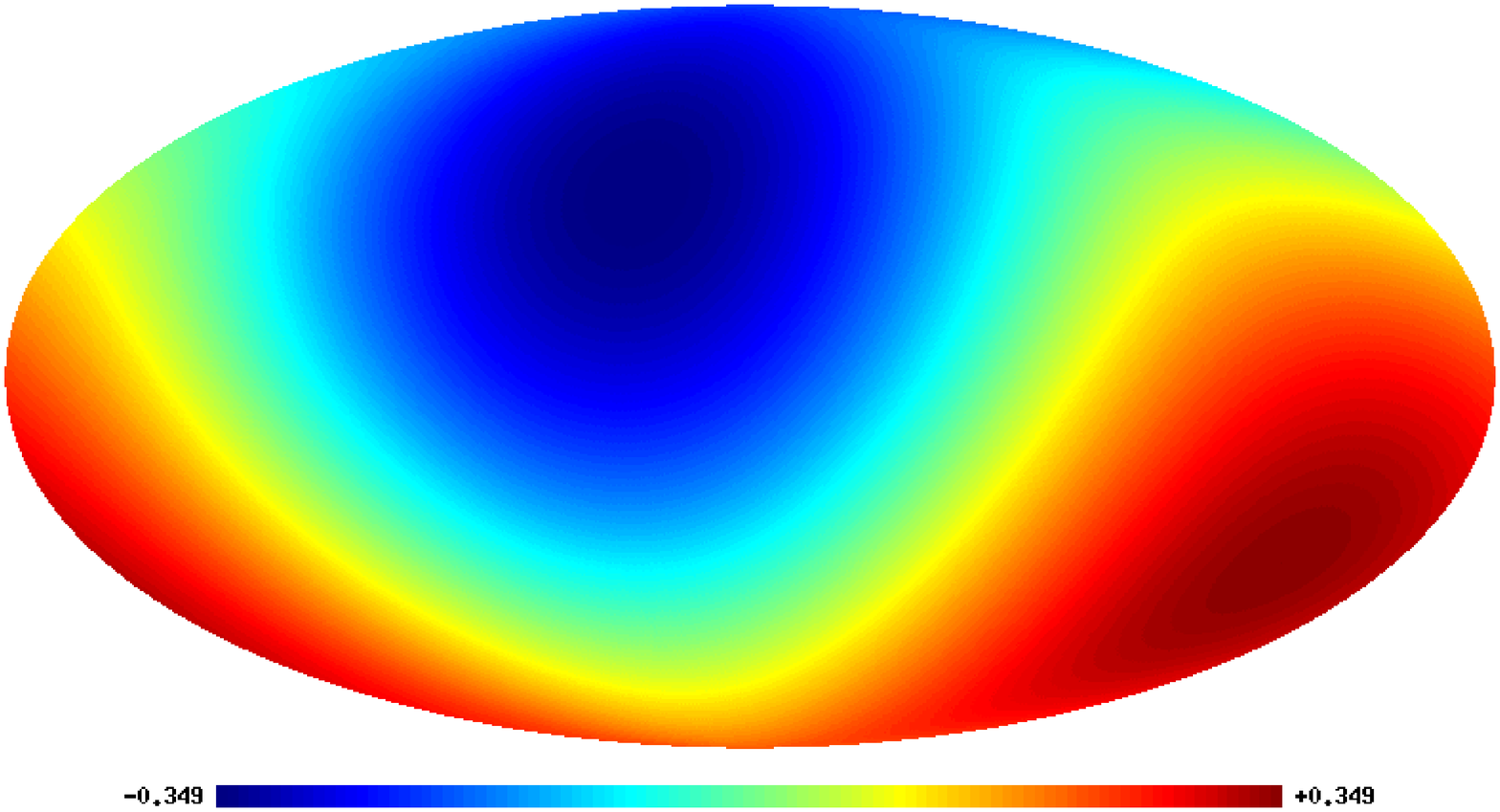}\includegraphics[width = 4cm]{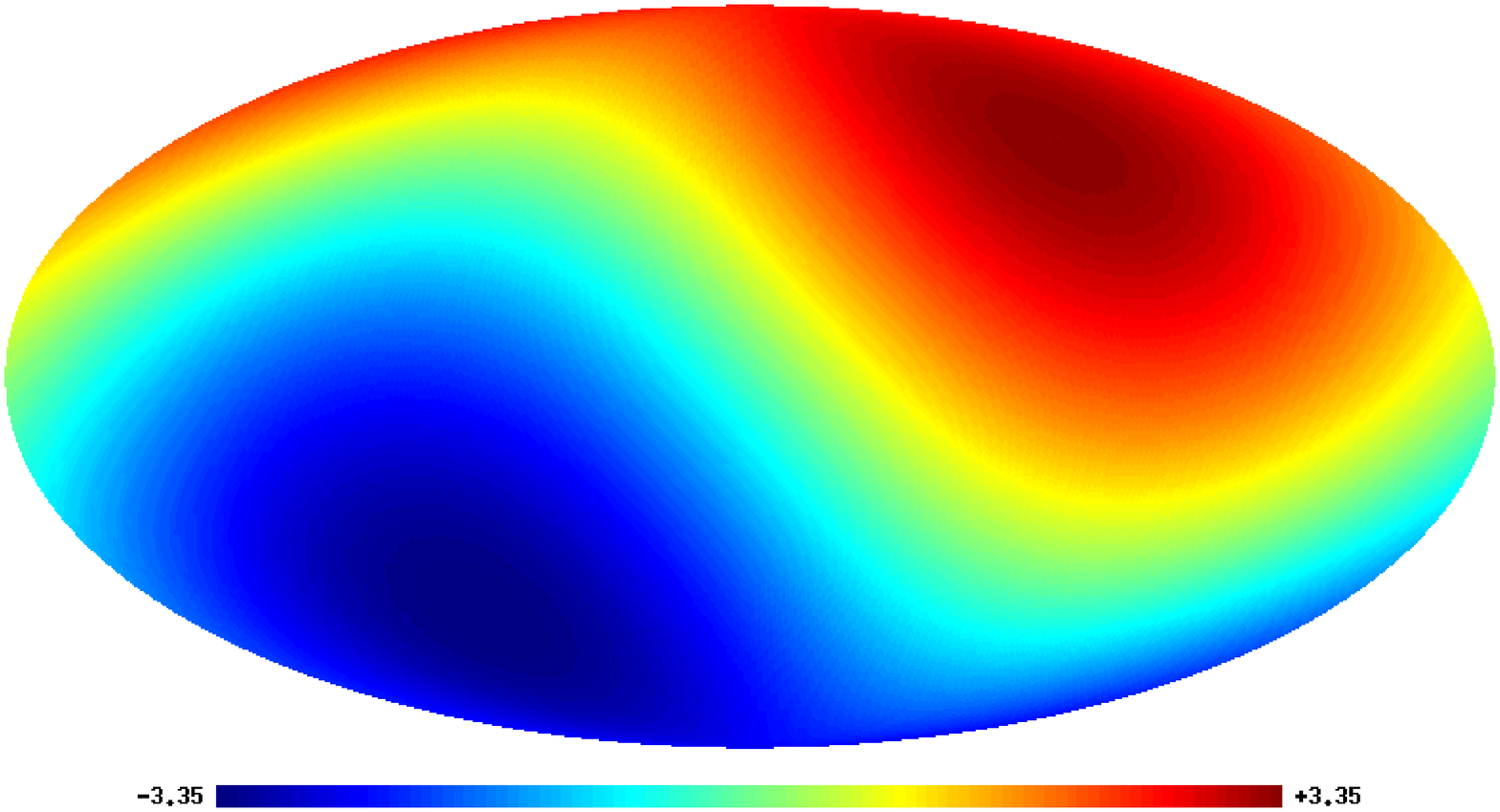}
\end{center}
\caption{Left Panel: The dipole component in the upper panel of Fig. \ref{figure2}. Right Panel: CMB kinematic dipole with the unit: mK.}\label{figure3}
\end{figure}

However, if we subtract this fit monopole from the $q_0$-map, the
residual is still significant, which has the similar amplitude
with the monopole component (see the middle panel in Fig.
\ref{figure2}). This shows that the anisotropy of $q_0$-map is
quite important, which is also the motivation of our discussion in
this paper. The lowest anisotropic component is the dipole. The
dipole is described by the amplitude $A_1$ and the direction
($\theta$, $\phi$) in Galactic coordinate system. For the
$q_0$-map, we get the fit dipole with the parameters ($A_1=0.349$,
$\theta=127.4^{\circ}$, $\phi=211.5^{\circ}$), which is plotted in
left panel of Fig. \ref{figure3}. We find that the amplitudes of
dipole and monopole are the same order. If subtracted both
monopole and dipole from the $q_0$-map, we find the residual
becomes very small, which is clearly shown in the lower panel in
Fig. \ref{figure2}. This implies that the main anisotropic
component in $q_0$-map is contributed by the dipole component.

Recently, a few puzzling large-scale cosmological observations
have been reported to challenge the standard model, which includes
the alignment of the low multipoles of CMB temperature and
polarization anisotropies
\cite{cmb_low_ls,cmb_low_ls2,cmb_low_ls3}, the parity asymmetry of
CMB power spectrum \cite{naselsky,naselsky2}, the large-scale
velocity flows \cite{bulk,bulk2} and the large scale alignment in
the QSO optical polarization data \cite{qso,qso2}. Especially, it
was noticed that these anomalies are all connected with the CMB
kinematic dipole in some sense \cite{hemi2,hemi22}. Even the
preferred axis of cosmic acceleration detected in
\cite{hemi1,hemi2,hemi22,cai,cai2} by using the hemisphere
comparison method is also claimed to align with the CMB kinematic
dipole. Here, we shall also compare the direction of the derived
dipole component with that of CMB kinematic dipole, which is
($A_1=3.35$mK, $\theta=41.74^{\circ}$, $\phi=263.99^{\circ}$)
\cite{dipole,dipole2} (see right panel in Fig. \ref{figure3}). {
Note that this dipole component was derived from the WMAP data,
and has been used by Planck mission for the calibration
\cite{planckdipole}.} The angle between these two dipoles is
$\alpha=97.6^{\circ}$. So we find that, instead of alignment,
these two dipoles are nearly perpendicular with each other.
Actually, the maximum axis reported by Antoniou and
Perivolaropoulos in \cite{hemi2,hemi22} is ($\theta=108^{\circ}$,
$\phi=129^{\circ}$) \footnote{Note that, the analysis does not
distinguish this direction and the opposite direction at
($\theta=72^{\circ}$, $\phi=309^{\circ}$).}. And the angle between
this direction and the CMB kinematic dipole direction is
$132.7^{\circ}$. The different between these two results may be
caused by the different direction resolutions of the methods.


\begin{figure}[t]
\begin{center}
\includegraphics[width = 4cm]{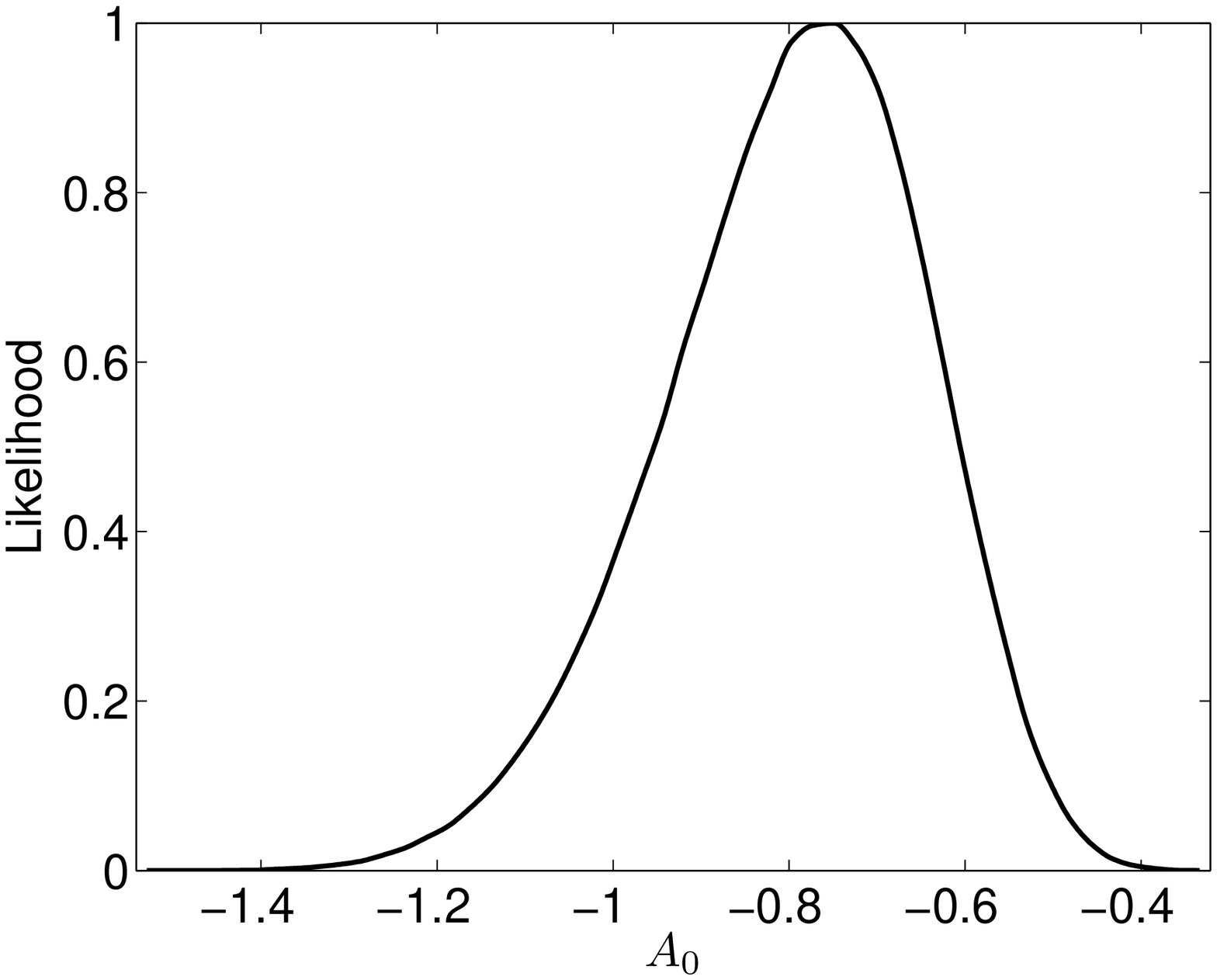}\includegraphics[width = 4cm]{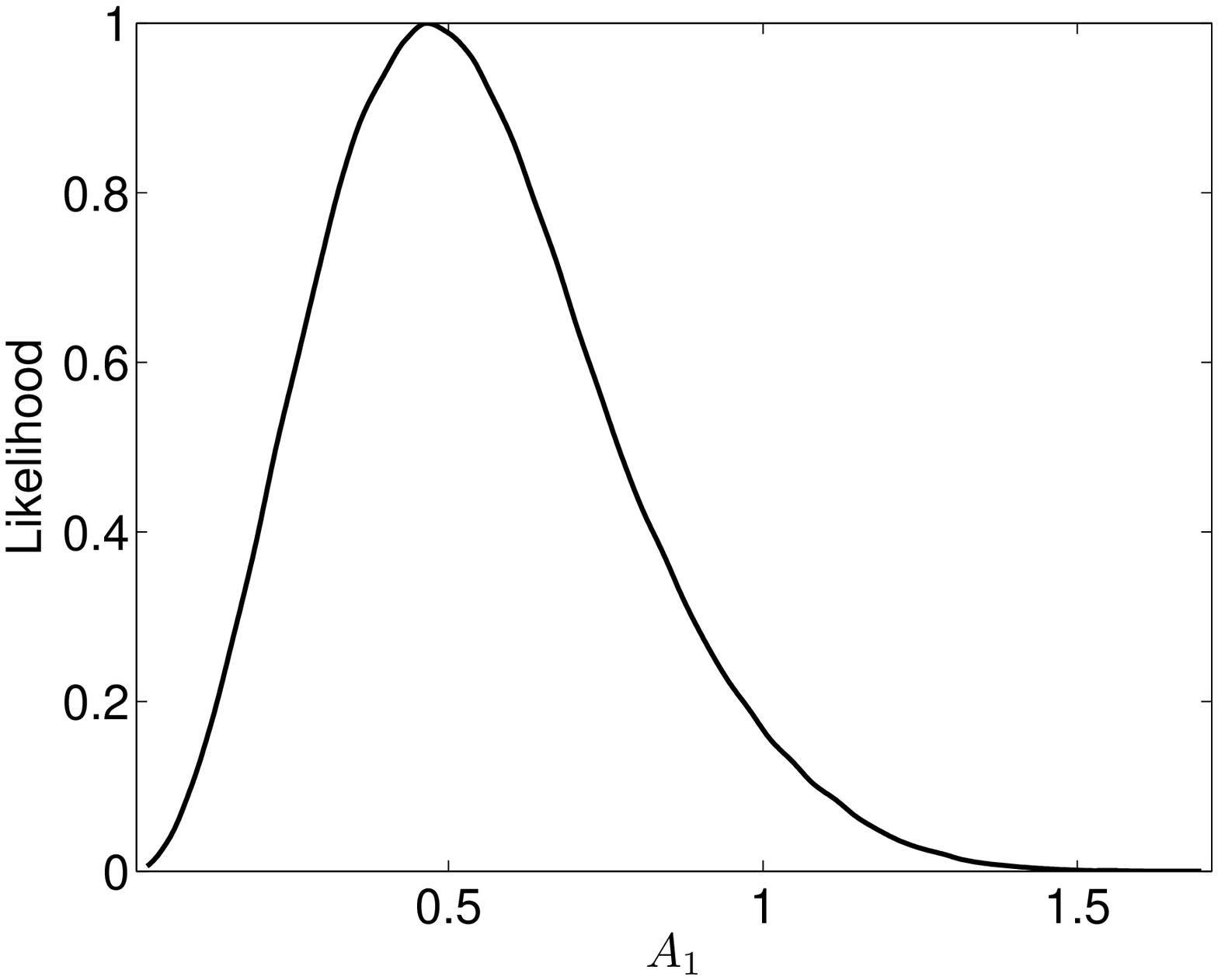}\\
\includegraphics[width = 4cm]{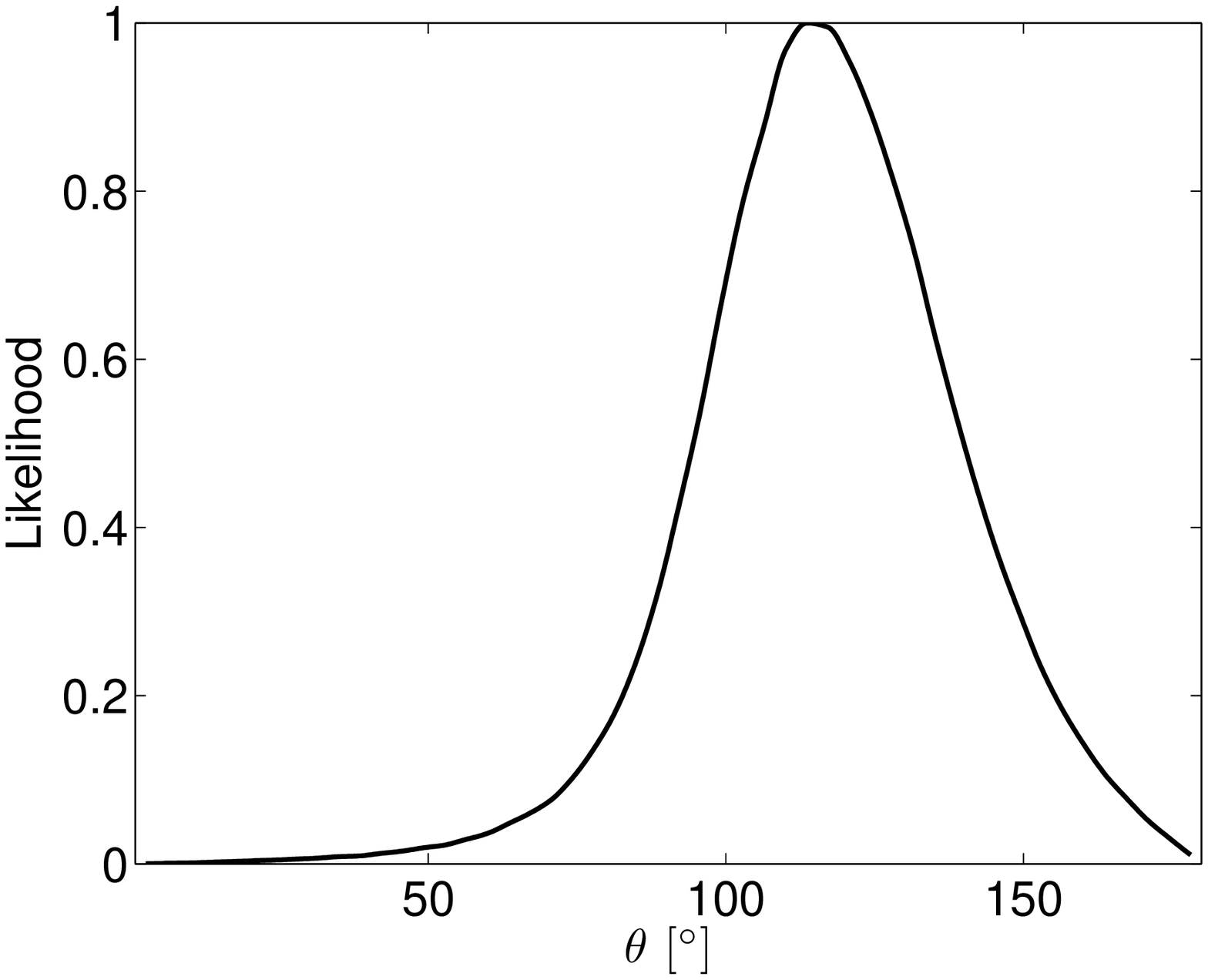}\includegraphics[width = 4cm]{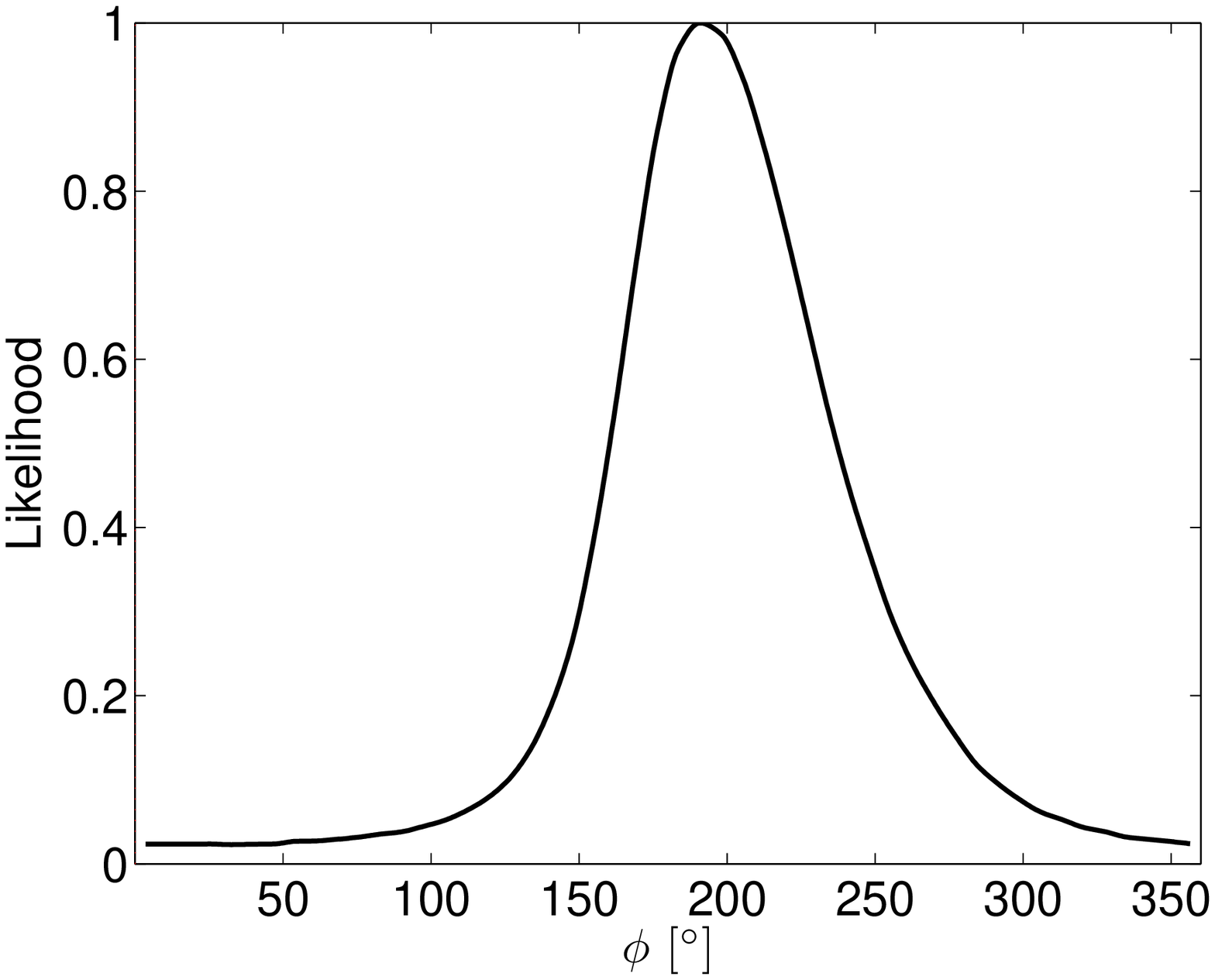}
\end{center}
\caption{The 1-dimensional likelihood functions for $A_0$, $A_1$, $\theta$ and $\phi$.}\label{figure5}
\end{figure}

Now, let us discuss the monopole and dipole components in the
general $q_0$-map, instead of those in the best-fit map. According
to the $q_0$ likelihood functions in the unmasked regions, we
generate $1,000,000$ random samples of $\{q_0^{2}, q_0^{3},
q_0^{4}, q_0^{9}, q_0^{10}, q_0^{11}\}$, where $q_0^i$ stands for
the $q_0$ in the $i$-th region. For each dataset, we derive the
corresponding monopole and dipole, and calculate the 1-dimensional
likelihood functions for $A_0$ and $A_1$ parameters, which are
shown in Fig. \ref{figure5}. We find that
$A_0=-0.750^{+0.122}_{-0.172}$, which is smaller than zero in more
than 5-$\sigma$ confident level. So the present Universe is in an
accelerating stage. Interesting enough, we also find that
$A_1=0.466^{+0.255}_{-0.205}$, i.e. the amplitude of dipole is
non-zero at more than 2-$\sigma$ level. So again, we find that the
dipole effect is quite significant.

\begin{figure}[t]
\begin{center}
\includegraphics[width = 4cm]{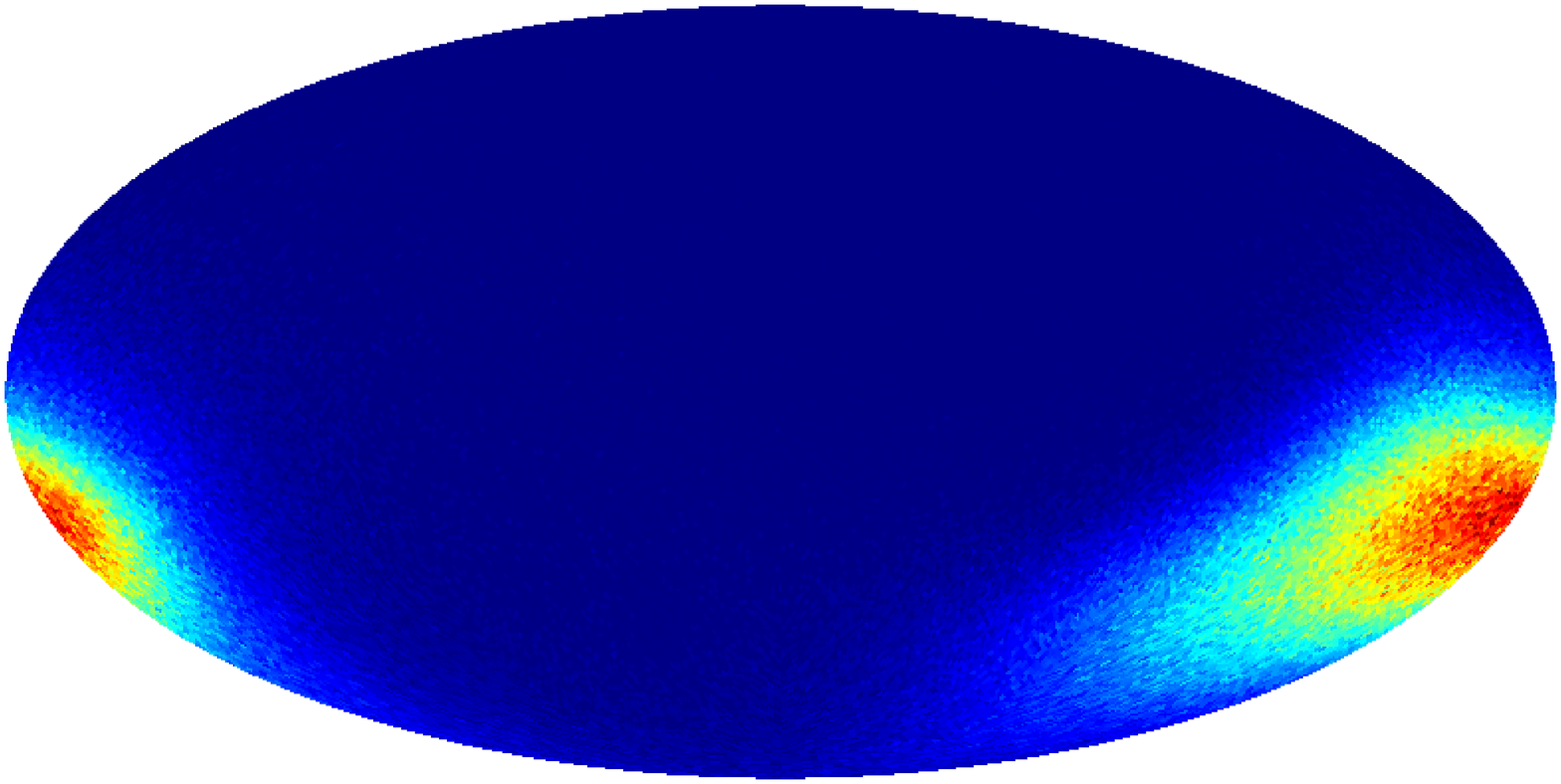}\includegraphics[width = 4cm]{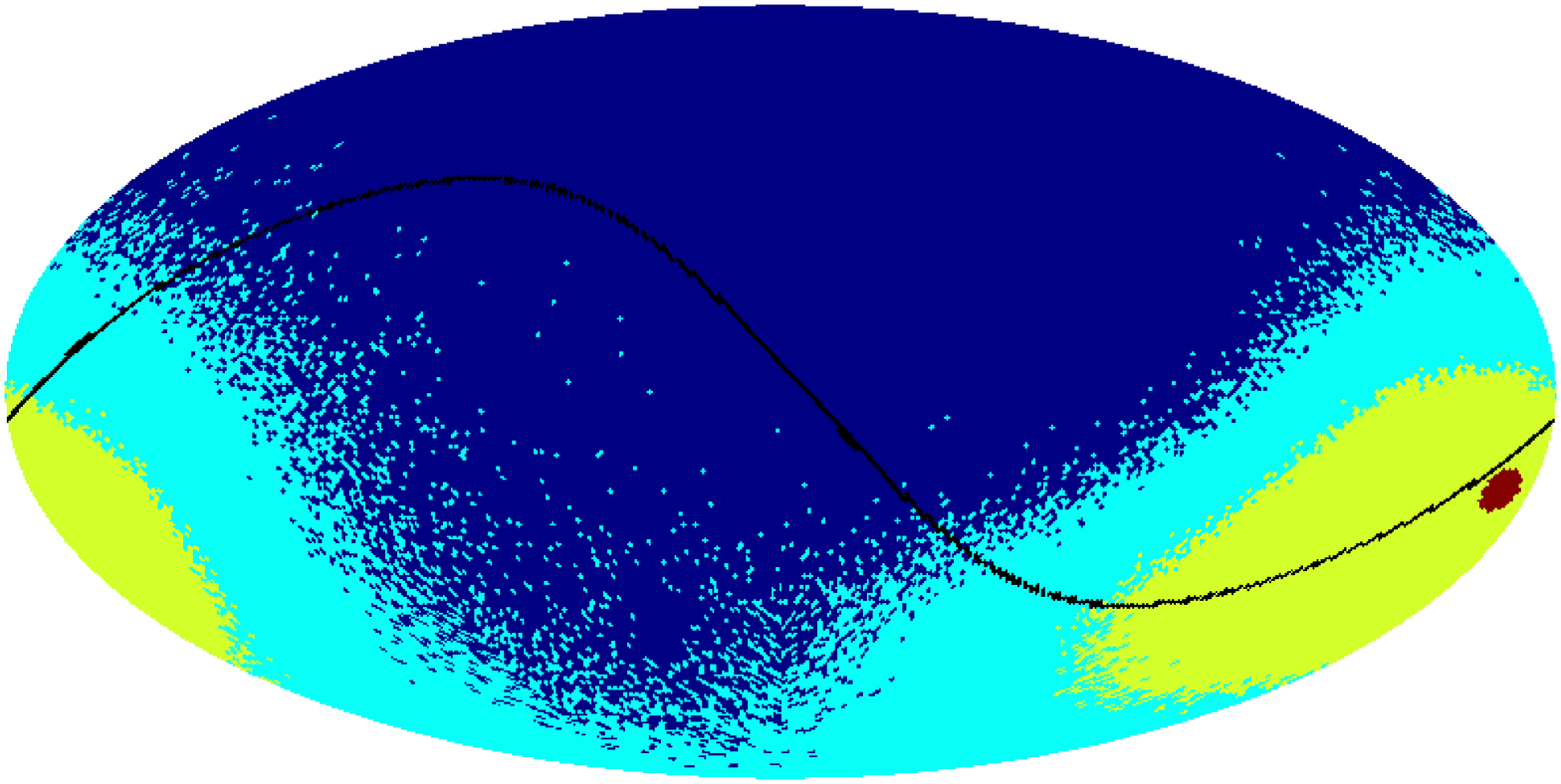}
\end{center}
\caption{Left Panel: The distribution of the dipole directions in Galactic coordinate system. Right Panel: The best-fit direction of dipole (red spot), 1-$\sigma$ region (yellow),
and 2-$\sigma$ region (cyan). The black line shows the plane, which is perpendicular to CMB kinematic dipole.}\label{figure4}
\end{figure}

In order to study the direction of the dipole component, in Fig.
\ref{figure4} we plot the likelihood distribution in 2-dimensional
space (left panel), and the corresponding best-fit value is
($\theta=108.8^{\circ}$, $\phi=187.0^{\circ}$). It is very
interesting to find that the angle between this best-fit dipole
and CMB kinematic dipole is $\alpha=95.7^{\circ}$. Again, we find
that these two dipoles are nearly perpendicular with each other.
In the right panel, we plot 1-$\sigma$ with yellow region and
2-$\sigma$ with cyan region. For comparison with CMB kinematic
dipole, in this figure we also plot the region with black line,
which is exactly perpendicular to CMB kinematic dipole. We find
this black line excellently crosses the centers of 1-$\sigma$ and
2-$\sigma$ regions, which is consistent with the above results.
The 1-dimensional likelihood functions for $\theta$ and $\phi$ are
also shown in Fig. \ref{figure5}, which correspond to the
constraints of $\theta=113.9^{+23.9}_{-17.2}$ and
$\phi=190.6^{+46.6}_{-30.3}$.

In order to quantify the perpendicular relation between these two
dipole directions $\hat{n}_1$ (the best-fit dipole of cosmic
acceleration) and $\hat{n}_2$ (CMB kinematic dipole), similar to
\cite{tegmark}, we can define the dot product $\hat{n}_1\cdot
\hat{n}_2$. Under the null hypothesis that these two dipoles are
statistically independent, with the unit vectors $\hat{n}_1$ and
$\hat{n}_2$ being independently drawn from a distribution where
all directions are equally likely. This means that the dot product
$|\hat{n}_1\cdot \hat{n}_2|$ is a uniformly distributed random
variable on the interval $[0, 1]$. By using the values of
$\hat{n}_1$ and $\hat{n}_2$, we get $|\hat{n}_1\cdot
\hat{n}_2|=0.10$, corresponding to a separation of $95.7^{\circ}$.
So a perpendicular this good happens by chance only once in
$1/0.10\simeq10$.

In the end of this section, although we will not detailedly study
the physical mechanism of the perpendicular relation in this
paper, we could provide some possible reasons for this coincidence
problem. Similar to other puzzles in the large-scale observations
\cite{cmb_low_ls,cmb_low_ls2,cmb_low_ls3,naselsky,naselsky2,bulk,bulk2,qso,qso2}
(see \cite{review} as a review), these anomalies may be the
indications of the non-trivial cosmic topology, such as the
Bianchi type models \cite{bianchi,bianchi2}. On the other hand, as
mentioned by some authors
\cite{fluctuation,fluctuation2,fluctuation3,hemi1}, these
coincidences may also hint some unsolved systematical errors in
observations or data analysis.


\section{Conclusions and discussions\label{concl}}

Recently, the anisotropy of cosmic acceleration has attracted
great attention, which may be caused by the non-trivial cosmic
topology or some residuals of observational errors. In this paper,
by using the Union2 SNIa dataset at different regions in the whole
sky, we investigated the dependence of cosmic acceleration on the
directions in the Galactic coordinate system, where the
deceleration parameter $q_0$ has been used as the diagnostic to
quantify the anisotropy level.

In the anisotropic $q_0$-maps, we find the significant dipole
effect with the amplitude $A_1=0.466^{+0.255}_{-0.205}$, which
deviates from zero at more than 2-$\sigma$ level. This study also
shows that the direction of the dipole trends to be perpendicular
to CMB kinematic dipole. The best-fit dipole direction is
($\theta=108.8^{\circ}$, $\phi=187.0^{\circ}$), and the angle
between this direction and that of CMB kinematic dipole is
$95.7^{\circ}$. We find the perpendicular relation between these
two dipoles is anomalous at the 1-in-10 level.

It is important to mention that as more and more supernovae data
will be released in the near future \cite{forcetask}, the much
more details of the anisotropy on cosmic acceleration could be
revealed (including the higher multipoles with $l\ge 2$ and the
dependence of redshift), which would be helpful to resolve the
origin of anisotropy on cosmic acceleration, and the association
with CMB kinematic dipole.

Recently, the new release of the Planck observations on the CMB
temperature anisotropy confirmed the alignment of the CMB
quadrupole and octupole. And this particular direction is nearly
aligned with CMB kinematic dipole direction \cite{planckproblem}.
At the same time, the discontinuous distribution of power in the
hemispheres on the sky was also been confirmed. All these show
that we have the evidence for a break in isotropy. In order to
solve these problems, a phenomenological dipole modulation may be
needed \cite{hu,planckproblem}. Since all these directional
anomalies, as well as the alignment problems of the cosmic
acceleration anisotropy discussed in this paper, the parity
asymmetry of CMB power spectrum \cite{naselsky}, the large-scale
velocity flows \cite{bulk} and the large scale alignment in the
QSO optical polarization data \cite{qso} are connected with the
CMB kinematic dipole and/or the ecliptic plane. We expect a single
dipole modulation mechanism could solve all these puzzles.

Several works have suggested that this kind of modulation could be
caused by the non-trivial topology of the Universe, such as the
anisotropic global Bianchi VIIh geometry \cite{plancktopology},
the Randers-Finsler geometry \cite{changzhe}, or the multi-stream
inflation \cite{wangyi}. However, if they have the cosmological
origin, it is very difficult to answer: Why the special direction
is related to the current motion direction of the Earth, i.e. the
CMB kinematic dipole. So, in our view, we would rather believe
that these problems should be caused by some unsolved systematical
errors in observations or data analysis. In any case, the much
more detailed investigations on this kind of directional anomalies
are necessary.


\section*{Acknowledgements}
The authors appreciate useful help from R. G. Cai and Z. L. Tuo,
and the helpful discussions with Q. G. Huang and Z. K. Guo. W.Z.
is supported by NSFC Nos. 11173021, 11075141 and project of
Knowledge Innovation Program of Chinese Academy of Science. P.W.
is supported by NSFC Nos. 11175093 and 11222545, Zhejiang
Provincial Natural Science Foundation of China under Grants No.
R6110518, the FANEDD under Grant No. 200922, and the NCET under
Grant No. 09-0144. Y.Z. is supported by NSFC Nos. 11073018, SRFDP
and CAS.



\begin{thebibliography}{35}

\bibitem{snia1}
S. Perlmutter, G. Aldering, G. Goldhaber, et al., Astrophys. J.  {\bf 517}, 565 (1999).

\bibitem{snia2}
A. G. Riess, A. V. Filippenko, P. Challis, et al., Astron. J. {\bf 116}, 1009 (1998).


\bibitem{peculiar}
C. Gordon, K. Land, and A. Slosar, Phys. Rev. Lett. {\bf 99},
081301 (2007).

\bibitem{peculiar2}
 C. Gordon, K. Land and A. Slosar, Mon. Not. Roy.
Astron. Soc. {\bf 387}, 371 (2008).

\bibitem{peculiar3}
 C. G. Tsagas, Mon. Not. Roy.
Astron. Soc. {\bf 405}, 503 (2010).

\bibitem{peculiar4}
 T. M. Davis, et al.,
Astrophys. J. {\bf 741}, 67 (2011).

\bibitem{fluctuation}
C. Bonvin, R. Durrer and M. A. Gasparini, Phys. Rev. D {\bf 73},
023523 (2006).

\bibitem{fluctuation2}
 M. Blomqvist, J. Enander and E. Mortsell, JCAP {\bf
10}, 018 (2010).

\bibitem{fluctuation3}
 S. A. Appleby1 and E. V. Linde, arXiv:1210.8221.

\bibitem{others}
S. Gupta, T. D. Saini and T. Laskar, Mon. Not. Roy. Astron. Soc.
{\bf 388}, 242 (2008).


\bibitem{others2}
 S. Gupta and T. D. Saini, Mon. Not. Roy.
Astron. Soc. {\bf 407}, 651 (2010).

\bibitem{others3}
L. Campanelli, P. Cea, G. L. Fogli and A. Marrone, Phys. Rev. D
{\bf 83}, 103503 (2011).

\bibitem{others4}
 J. Colin, R. Mohayaee, S. Sarkar and A.
Shafieloo, Mon. Not. Roy. Astron. Soc. {\bf 414}, 264 (2011).

\bibitem{hemi1}
D. J. Schwarz and B. Weinhorst, Astron. Astrophys. {\bf 474},
717 (2007).

\bibitem{hemi2}
I. Antoniou and L. Perivolaropoulos, JCAP {\bf 1012}, 012 (2010);

\bibitem{hemi22}
A. Mariano and L. Perivolaropoulos, arXiv:1206.4055.

\bibitem{cai}
R. G. Cai and Z. L. Tuo, JCAP {\bf 1202}, 004 (2012).

\bibitem{kalus}
B. Kalus, D. J. Schwarz, M. Seikel and A. Wiegand, Astron.
Astrophys. {553}, A56 (2013).

\bibitem{cai2}
 R. G. Cai, Y. Z. Ma, B. Tang and Z. L. Tuo, arXiv:1303.0961.

\bibitem{vector}
C. Armendariz-Picon, JCAP {\bf 0407}, 007 (2004).

\bibitem{ym}
W. Zhao and Y. Zhang, Class. Quant. Grav. {\bf 23}, 3405 (2006).

\bibitem{ym2}
W. Zhao and Y. Zhang, Phys. Lett. B {\bf 640}, 69 (2006).


\bibitem{ym3}
Y. Zhang, T. Y. Xia, W. Zhao, Class. Quant. Grav. {\bf 24}, 3309
(2007).

\bibitem{vector2}
T. S. Koivisto and D. F. Mota, JCAP {\bf 0808}, 021 (2008).

\bibitem{ani}
T. Koivisto and D. F. Mota, \apj {\bf 679}, 1 (2008).


\bibitem{bianchi}
T. R. Jaffe, S. Hervik, A. J. Banday and K. M. Gorski, Astrophys.
J. {\bf 644}, 701 (2006).


\bibitem{bianchi2}
A. Pontzen and A. Challinor, Mon. Not. Roy. Astron. Soc. {\bf
380}, 1387 (2007).

\bibitem{topology}
P. Bielewicz and A. Riazuelo, Mon. Not. Roy. Astron. Soc. {\bf 396}, 609 (2009).

\bibitem{perturbation}
C. Armendariz-Picon, JCAP {\bf 0709}, 014 (2007).

\bibitem{perturbation2}
A. R. Pullen and M. Kamionkowski, Phys. Rev. D {\bf 76}, 103529
(2007).

\bibitem{magnetic}
M. Giovannini and M. Shaposhnikov, Phys. Rev. D {\bf 62}, 103512
(2000).

\bibitem{magnetic2}
T. Kahniashvili, G. Lavrelashvili and B. Ratra, Phys. Rev. D {\bf
78}, 063012 (2008).

\bibitem{magnetic3}
 J. Kim and P. Naselsky, JCAP {\bf 0907}, 041
(2009).

\bibitem{principle}
T. S. Kolatt and O. Lahav, \mnras ~{\bf 323}, 859 (2001).

\bibitem{principle2}
T. Clifton, P. G. Ferreira and K. Land, \prl~ {\bf 101}, 131302
(2008).

\bibitem{principle3}
 T. J. Zhang, H. Wang and C. Ma, arXiv:1210.1775.

\bibitem{union2}
R. Amanullah,  et al., Astrophys. J. {\bf 716}, 712 (2010).

\bibitem{cfa}
M. Hicken, et al., Astrophys. J. {\bf 700}, 1097 (2009).

\bibitem{sdss}
J. A. Holtzman, et al., Astron. J. {\bf 136}, 2306 (2008).

\bibitem{healpix}
K. M. Gorski, E. Hivon, A. J. Banday, B. D. Wandelt, F. K. Hansen, M. Reinecke and M. Bartelman, Astrophys. J. {\bf 622}, 759 (2005).

\bibitem{chi2}
S. Nesseris and L. Perivolaropoulos, Phys. Rev. D {\bf 72}, 123519 (2005).

\bibitem{other_results}
Y. Gong and A. Wang,  Phys. Rev. D {\bf  75}, 043520 (2007).

\bibitem{other_results2}
P. Wu and H. Yu,  JCAP {\bf 0802}, 019 (2008).

\bibitem{cmb_low_ls}
K. Land and J. Magueijo, Phys. Rev. Lett. {\bf 95}, 071301 (2005).

\bibitem{cmb_low_ls2}
M. Frommert and T. A. Ensslin, Mon. Not. Roy. Astron. Soc. {\bf
403}, 1739 (2010).

\bibitem{cmb_low_ls3}
C. L. Bennett, et al., \apjs ~{\bf 192}, 17 (2011).


\bibitem{naselsky}
J. Kim and P. Naselsky, \apjl ~{\bf 714}, L265 (2010)

\bibitem{naselsky2}
P. Naselsky, W. Zhao, J. Kim and S. Chen, Astrophys. J. {\bf 749},
31 (2012).

\bibitem{bulk}
R. Watkins, H. A. Feldman and M. J. Hudson, Mon. Not. Roy. Astron.
Soc. {\bf 392}, 743 (2009).

\bibitem{bulk2}
H. A. Feldman, R. Watkins and M. J. Hudson, Mon. Not. Roy. Astron.
Soc. {\bf 407}, 2328 (2010).


\bibitem{qso}
D. Hutsemekers, R. Cabanac, H. Lamy and D. Sluse, Astron. Astrophys. {\bf 441}, 915 (2005);

\bibitem{qso2}
D. Hutsemekers, A. Payez, R. Cabanac, H. Lamy, D. Sluse, B.
Borguet and J. R. Cudell, arXiv:0809.3088.


\bibitem{dipole}
N. Jarosik, et al., \apjs~ {\bf 192}, 14 (2011).

\bibitem{dipole2}
G. Hinshaw, et al., \apjs~ {180}, 225 (2009).

\bibitem{planckdipole}
Planck Collaboration, arXiv:1303.5069.

\bibitem{tegmark}
A. de Oliveira-Costa, M. Tegmark, M. Zaldarriaga and A. Hamilton, Phys. Rev. D {\bf 69}, 063516 (2004).


\bibitem{review}
L. Perivolaropoulos, arXiv:1104.0539.

\bibitem{forcetask}
A. Albrecht, et al., arXiv:astro-ph/0609591.

\bibitem{planckproblem}
Planck Collaboration, arXiv:1303.5083.

\bibitem{hu}
C. Gordon, W. Hu, D. Huterer and T. Crawford, \prd {\bf 72},
103002 (2005).

\bibitem{plancktopology}
Planck Collaboration, arXiv:1303.5086.

\bibitem{changzhe}
Z. Chang and S. Wang, arXiv:1303.6058.

\bibitem{wangyi}
Y. Wang, arXiv:1304.0599.

\end{thebibliography}
\end{document}